\documentclass[conference]{IEEEtran}
\IEEEoverridecommandlockouts
\usepackage{cite}
\usepackage{amsmath,amssymb,amsfonts}
\usepackage{algorithmic}
\usepackage{graphicx}
\usepackage{caption}
\usepackage{subcaption}
\usepackage{textcomp}
\usepackage{xcolor}
\usepackage{comment}
\usepackage{svg}
\usepackage{algorithm}
\usepackage{array}
\usepackage{multirow}
\usepackage[left=0.673in,right=0.673in,top=0.7in,bottom=1.013in]{geometry}
\newcolumntype{L}[1]{>{\raggedright\let\newline\\\arraybackslash\hspace{0pt}}m{#1}}
\def\BibTeX{{\rm B\kern-.05em{\sc i\kern-.025em b}\kern-.08em
    T\kern-.1667em\lower.7ex\hbox{E}\kern-.125emX}}

\allowdisplaybreaks

\begin{document}

\title{Energy Management for Prepaid Customers: \\A Linear Optimization Approach
\thanks{\noindent The authors acknowledge support from the the U.S. National Science Foundation under Award Number ECCS-2045860 and the George Bunn Distinguished Graduate Fellowship by the University of Wisconsin-Madison.}
}

\author{\IEEEauthorblockN{Maitreyee Marathe and Line A. Roald}
\IEEEauthorblockA{\textit{Dept. of Electrical and Computer Engineering,}
\textit{University of Wisconsin-Madison,}
Madison, USA \\
\{mmarathe, roald\}@wisc.edu
}
}

\maketitle

\begin{abstract}
With increasing energy prices, low income households are known to forego or minimize the use of electricity 
to save on energy costs. If a household is on a prepaid electricity program, it can be automatically and immediately disconnected from service if there is no balance in its prepaid account. Such households need to actively ration the amount of energy they use by deciding which appliances to use and for how long. We present a tool that helps households extend the availability of their critical appliances by limiting the use of discretionary ones, and 
prevent disconnections. The proposed method is based on a linear optimization problem that only uses average power demand as an input and can be solved to optimality using a simple greedy approach. We compare the model with two mixed-integer linear programming models that require more detailed demand forecasts and optimization solvers for implementation.
In a numerical case study based on real household data, we assess the performance of the different models under different accuracy and granularity of demand forecasts. Our results show that our proposed linear model is much simpler to implement, while providing similar performance under realistic circumstances.
\end{abstract}

\section{Introduction}
Unaffordable electricity poses a major hurdle for many  households to meet their energy needs. Low-income households are known to forego energy use to pay for other critical needs, referred to as `energy limiting behavior' \cite{cong2022unveiling}. Rationing energy, i.e., choosing which electric appliances (loads) to power and which ones to curtail, is an every day reality for such households. 
This can lead to unsafe practices such as allowing unhealthy indoor temperatures to avoid switching on the air conditioner \cite{doremus2022sweating}. Households that are on prepaid plans, i.e., who pay for their electricity use a-priori by adding money to a prepaid wallet (which we will refer to as `recharging' their wallet), are particularly vulnerable because they can be automatically and immediately disconnected from supply if their wallet balance reaches zero. There are several million prepaid electricity customers across North America and Europe \cite{noauthor_smart_2017}, \cite{2022_smartenergyinternational}. Prepaid metering is a popular choice among low-income customers because it offers flexibility in payments and does not need credit checks. However, unanticipated disconnections and high reconnection fees can cause significant inconvenience and may even be unsafe during extreme weather \cite{2020_nclc}. Therefore, low-income households, and particularly those on prepaid metering, need a framework for effectively managing energy use within a limited budget. 

As discussed in \cite{beaudin2015home}, home energy management systems generally aim to optimize one or more of the following objectives: \textit{cost}, \textit{well-being}, \textit{emissions}, and \textit{load profile}. In the context of energy rationing, we aim to minimize user inconvenience (maximize \textit{well-being}) within a fixed energy budget (a constraint on \textit{cost}). Further, \cite{beaudin2015home} classify inconvenience modeling into two types: ``inconvenience due to timing" (e.g., shifting a load from the morning to afternoon), and ``inconvenience due to undesirable energy states". The latter is most closely aligned with our setting, since low-income families 
with a limited budget
may 
enter the undesirable energy state
where they have to forego using a load in order to preserve wallet balance for using more critical loads in the future. 

Very few studies investigate energy management methods for prepaid customers. The method in \cite{souza2020automatic} uses load disaggregation techniques and alerts the user to switch off a load to preserve wallet balance once it exceeds a target consumption. To perform the disaggregation, additional hardware or high-speed internet connectivity with a server may be needed. More traditional home energy management systems (HEMS) \cite{shareef2018review} generate specific load schedules to switch loads ON/OFF. This can be viewed as intrusive, and if automated, it needs expensive in-home hardware such as smart switches for actuating loads. 
Furthermore, to be effective, such HEMS typically need forecasts of power demand of each load at each timestep in a day. 
Obtaining highly accurate, granular forecasts can be challenging because of the volatile nature of household-level electricity use, which depends on various behavioral and environmental factors \cite{li2021short}.
Furthermore, imperfect and uncertain information can significantly impact the performance of HEMS \cite{beaudin2015home}, \cite{blonsky2022home}, making it necessary to analyse sensitivity of energy management methods to imperfect information.
Overall, energy management methods that use lower granularity forecasts (e.g., by aggregating loads or averaging in time) may lend themselves better to practical implementation than very detailed models. 

\subsection{Threshold-based energy management}

\begin{figure}
\centering
\includegraphics[width= 0.85\columnwidth]{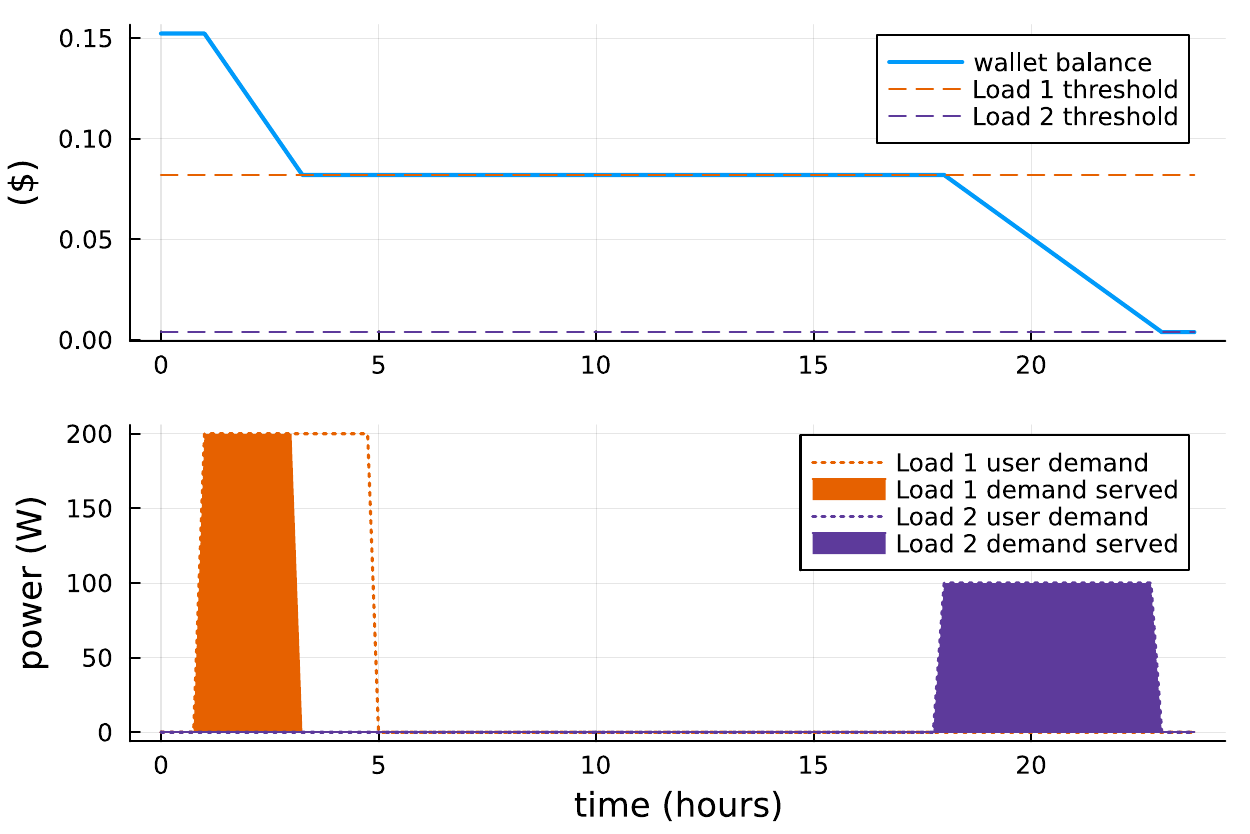}
\caption{\small Illustration of threshold-based energy management with wallet balance and load thresholds (top figure) and load demand (bottom figure). Load demand is served when the wallet balance is greater than the corresponding threshold. Note that load 1 is served only partially, while load 2 is served fully. 
}
\vspace{-6mm}
\label{fig:threshold-based-energy-mgmt-illustration}
\end{figure}

To support effective home energy rationing while reducing the need for communication and computation, \cite{marathe2023optimal} proposed a \emph{threshold-based energy management framework} which is adapted from a load-control scheme for DC microgrids \cite{manur2020distributed}. 
Instead of generating switching schedules for loads, threshold-based energy management
uses enable/disable thresholds for each load, expressed in terms of the prepaid wallet balance. If the balance drops below this threshold, the load is disabled, i.e., the user is notified that they should switch off the load to preserve wallet balance for future use. This is illustrated in Figure \ref{fig:threshold-based-energy-mgmt-illustration}.
To identify optimal daily load thresholds, \cite{marathe2023optimal} 
uses rolling-horizon mixed-integer optimization with forecasts of load demand for every $15\text{ } \mathrm{min}$. However, while the thresholds optimization requires load demand forecasts, real-time measurements of load demand are not necessary once the thresholds are determined.

\subsection{Contributions}
We make the following contributions beyond prior work:

First, we present a new threshold-based energy management method, which has two main features that distinguish it from the model in \cite{marathe2023optimal}: (1) It only requires information regarding the average power demand per day, as compared to detailed power demand forecasts at each $15\text{ }\mathrm{min}$ timestep. In particular, it does not require information about when certain loads will be used during the day. 
(2) The resulting optimization problem is a linear program (LP), which can be solved to optimality using a greedy approach, thus eliminating the need for a solver.

Next, we compare the proposed model to two benchmark models, the threshold-based model from \cite{marathe2023optimal} and a more traditional home energy management system that generates a switching schedule for each load at each timestep. 
We first qualitatively compare the practicality of implementing and using the different methods, including the computation, demand forecast information, and communication requirements. 
Furthermore, we perform a quantitative comparison where we run the three models on a case study based on real load data from Pecan Street Dataport \cite{noauthor_dataport_nodate}. We analyze the performance of the three models under different levels of accuracy and granularity of demand forecast information.

The results of our comparison show that the proposed method has a similar quantitative performance as the two benchmark models, i.e., it is able to help customers effectively manage their electric loads. Furthermore, it has significantly lower requirements for demand forecasts and communication, can be implemented using simpler hardware without optimization solvers, making it a much more viable choice for low-income customers. The model is made publicly available through a GitHub repository \cite{github-repo}.

\subsection{Paper organization}
The paper is organized as follows: Section \ref{section:models} describes the mathematical optimization formulations for the three models. Section \ref{section:hardware-communication-requirements} compares computation, forecast information, and communication requirements for implementing the models. Section \ref{section:case-studies} presents case studies comparing the performance of the models using real-world energy usage data, and this is followed by a brief concluding section.
\section{Model Formulations}
\label{section:models}

\begin{table}[]
    \caption{Nomenclature}
    \setlength{\tabcolsep}{3pt}
    \begin{tabular}{p{0.19\columnwidth} p{0.5\columnwidth} p{0.21\columnwidth}}
        \multicolumn{2}{l}{\textbf{Parameters}} & \textbf{Model}\\
        $\mathcal{K}$ & Set of all loads & {\scriptsize AFG, DFM, OBM}\\
$\mathcal{T}$ & Set of all time steps in the horizon & {\scriptsize DFM, OBM}\\
$\Delta T \in \mathbb{R}_{>0}$ & Length of time step in  hours & {\scriptsize DFM, OBM}\\
$\mathcal{D}$ & Set of all days in the horizon & {\scriptsize AFG, DFM}\\
$m \in \mathbb{R}_{<0}$ & Large magnitude constant  & {\scriptsize DFM}\\
$M \in \mathbb{R}_{>0}$ & Large magnitude constant  & {\scriptsize DFM}\\
$\epsilon \in \mathbb{R}_{>0}$ & Small magnitude constant & {\scriptsize DFM}\\
$\alpha \in \mathbb{R}_{>0}$ & Electricity rate in \$/Wh & {\scriptsize AFG, DFM, OBM}\\
$Z \in \mathbb{R}_{\geq 0}$ & Initial real wallet balance in \$ & {\scriptsize AFG, DFM, OBM}\\
$\gamma_k \in \mathbb{R}_{>0}$ & Priority factor for load $k$  & {\scriptsize AFG, DFM, OBM}\\
$P_{k,t} \in \mathbb{R}_{\geq 0}$ & Demand in W for load $k$ at time $t$ & {\scriptsize DFM, OBM}\\
$\bar{P}_{k,d} \in \mathbb{R}_{\geq 0}$ & Average demand in W for load $k$ on day $d$ & {\scriptsize AFG}\\
$d_{k,t} \in \{0,1\}$ & Indicator parameter of demand & {\scriptsize DFM, OBM}\\
& $d_{k,t} = 1$ if $P_{k,t} > 0$, and $0$ otherwise & \\
$X_d \in \mathbb{R}_{\geq 0}$ & Virtual wallet recharge on day $d$ in \$ & {\scriptsize DFM}\\
$s_{k,d}^{max} \in \mathbb{R}_{\geq 0}$ & Upper bound on enable duration of load $k$ on day $d$ in hours & {\scriptsize AFG}\\[+3pt]
        \multicolumn{2}{l}{\textbf{Variables}} & \\
        $\mathbf{z}_t \in \mathbb{R}$ & Real wallet balance at time $t$ in \$ & {\scriptsize DFM}\\
$\mathbf{u}^z_{k,t} \in \{0,1\}$ & Real enable signal for load $k$ at time $t$ & {\scriptsize DFM}\\
$\mathbf{x}_t \in \mathbb{R}$ & Virtual wallet balance at time $t$ in \$ & {\scriptsize DFM}\\
$\mathbf{x}_{k,d} \in \mathbb{R}$ & Threshold for load $k$ on day $d$ in \$ & {\scriptsize AFG, DFM}\\
$\mathbf{u}^x_{k,t} \in \{0,1\}$ & Virtual enable signal for load $k$ at time $t$ & {\scriptsize DFM}\\ 
$\mathbf{a}_{k,t} \in \{0,1\}$ & Actuation state of load $k$ at time $t$ & {\scriptsize DFM, OBM}\\
$\mathbf{s}_{k,d} \in \mathbb{R}_{\geq 0}$ & Enable duration of load $k$ on day $d$ in hours & {\scriptsize AFG}\\
$\mathbf{X}_{d} \in \mathbb{R}_{\geq 0}$ & Virtual wallet recharge on day $d$ in \$ & {\scriptsize AFG}
    \end{tabular}
    \label{tab:nomenclature}
\vspace{-4mm}
\end{table}

This section presents the mathematical optimization formulations of the three models. A nomenclature with the notation used across models is given in Table \ref{tab:nomenclature}.

The objective of all models is to maximize the value that a user can get from a limited amount of initial wallet balance.
Concretely, we aim to maximize the fraction of time for which a load was ON when demanded. We refer to this as the \emph{service factor} $\mathrm{SF}_k$ for load $k$. For example, a lamp demanded for 2 hours and served for 1 hour has a service factor of $50\%$. Since some loads are more critical than others, we assign a \textit{priority factor} $\gamma_k$ to each load $k$, where a higher $\gamma$ indicates a more important load. 
Finally, we define the priority service factor (PSF) as the priority factor weighted sum of service factors, $\mathrm{PSF} = \sum\limits_{k\in\mathcal{K}}\gamma_{k}\mathrm{SF}_k$. PSF is a linear measure of user convenience and well-being, and all the models presented below seek to maximize PSF.

\subsection{Detailed Forecast MILP (DFM) model}
We first summarize the DFM model from \cite{marathe2023optimal},
which we will use as a benchmark.

\subsubsection{Modeling considerations}
\begin{figure}
\centering
\includegraphics[width= 0.62\columnwidth]{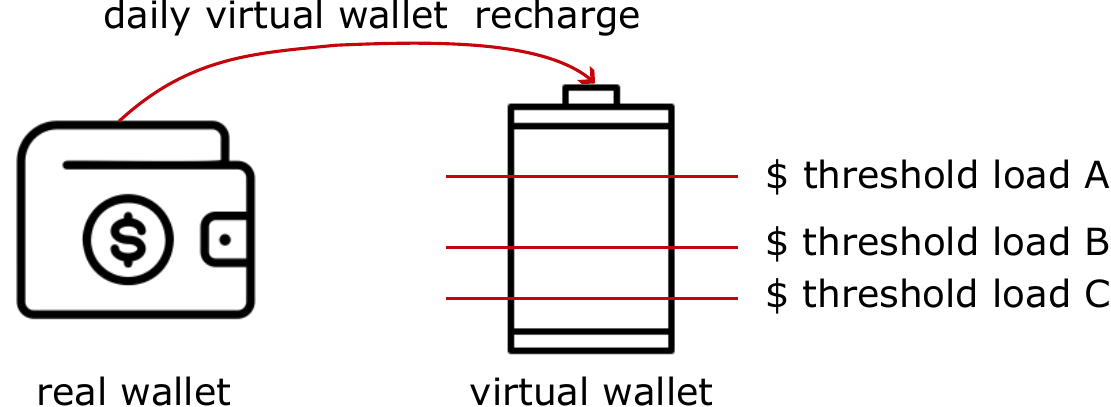}
\caption{\small Illustration of virtual wallet recharge} 
\vspace{-10pt}
\label{fig:virtual-wallet}
\end{figure}

The DFM model uses demand forecasts $P_{k,t}$ per load $k$ at each timestep $t$ and
uses rolling-horizon optimization to determine the optimal thresholds $\mathbf{x}_{k,d}$, per load $k$ per day $d$. 
The load thresholds are expressed in terms of prepaid wallet balance (i.e., in \$), as the prepaid wallet balance is a measure of how much energy is available (similar to the battery state of charge in the case of DC microgrid control \cite{manur2020distributed}).
However, the prepaid wallet may be recharged at infrequent and possibly irregular intervals. To avoid the balance being used too quickly, \cite{marathe2023optimal} defines a \textit{virtual wallet}. This wallet is recharged with regular amounts from the actual prepaid wallet, which we refer to as the \textit{real wallet}, and thresholds are defined in terms of the virtual wallet balances. This setup is illustrated in Figure \ref{fig:virtual-wallet}. 
The DFM model based on \cite{marathe2023optimal} defines the the virtual wallet recharge $X_d$ as 
the initial real wallet balance $Z$ divided by the number of days in the optimization horizon. 

\subsubsection{Mathematical formulation}
The mathematical formulation of the DFM problem is given as follows,
\begin{subequations}
    \begin{align}
        &\max_{\substack{\mathbf{x}_{k,d}, \mathbf{x}_t, \mathbf{z}_t, \mathbf{u}_{k,t}^x, \mathbf{u}^z_{k,t}, \mathbf{a}_{k,t}}} &&\mathrm{PSF} = \sum\limits_k \gamma_k \frac{\sum\limits_t \mathbf{a}_{k,t}}{\sum\limits_t d_{k,t}}
        \label{eq:OBM-objective}
    \end{align}
    \begin{align}
        &\text{~s.t.}&&\mathbf{z}_{t} = \mathbf{z}_{t-1} + Z - \alpha\Delta T \sum_{k}(P_{k,t-1}\mathbf{a}_{k,t-1}) \quad \forall t \in \mathcal{T}
        \label{eq:DFM-real-recharge} \\
        &&&m\mathbf{u}^z_{k,t}\leq -\mathbf{z}_t \quad\quad\quad\quad\quad\quad\quad~ \forall k \in \mathcal{K}, \forall t \in \mathcal{T}
        \label{eq:DFM-real-enable}\\
        &&&(M+\epsilon)(1-\mathbf{u}^z_{k,t})\geq \epsilon - \mathbf{z}_t \quad~ \forall k \in \mathcal{K}, \forall t \in \mathcal{T}
        \label{eq:DFM-real-disable} \\
        &&&\mathbf{x}_t = \mathbf{x}_{t-1} + X_d - \alpha\Delta T\sum_{k}(P_{k,t-1}\mathbf{a}_{k,t-1}) \quad \forall t \in \mathcal{T}
        \label{eq:DFM-virtual-recharge} \\
        &&&\mathbf{x}_t - \mathbf{x}_{k,d} + \epsilon \leq (M+\epsilon)\mathbf{u}^x_{k,t} \quad \forall k \in \mathcal{K}, \forall t \in \mathcal{T}
        \label{eq:DFM-virtual-enable}\\
        &&&\mathbf{x}_t - \mathbf{x}_{k,d} \geq m(1-\mathbf{u}^x_{k,t}) \quad\quad~~ \forall k \in \mathcal{K}, \forall t \in \mathcal{T}
        \label{eq:DFM-virtual-disable} \\
        &&&\mathbf{a}_{k,t}\!\leq\! d_{k,t}\mathbf{u}^x_{k,t}, ~~ \mathbf{a}_{k,t}\!\leq\!\mathbf{u}^z_{k,t}, ~~ \mathbf{a}_{k,t} \!\leq\!\mathbf{u}^z_{k,t+1}  \nonumber \\
        &&& \qquad\qquad\qquad\qquad\qquad\qquad~~~\forall k\!\in\!\mathcal{K}, \forall t\!\in\!\mathcal{T} \label{eq:actuation-1} \\
        &&&d_{k,t}\mathbf{u}^x_{k,t} + \mathbf{u}^z_{k,t} + \mathbf{u}^z_{k,t+1} \leq 2 + \mathbf{a}_{k,t} \nonumber \\&&& \qquad\qquad\qquad\qquad\qquad\qquad~~~\forall k \in \mathcal{K}, \forall t \in \mathcal{T}
        \label{eq:actuation-on}
    \end{align}
\end{subequations}
Here, the objective function is to maximize the PSF as given in (\ref{eq:OBM-objective}).
Constraints (\ref{eq:DFM-real-recharge}) ensure that the real wallet balance $\mathbf{z}_{t}$ is updated at each time step $t$; the initial real wallet balance $Z$ is included in the constraint corresponding to the first time step ($t=1$). Constraints (\ref{eq:DFM-real-enable}), (\ref{eq:DFM-real-disable}) ensure that the real enable signal $\mathbf{u}_{k,t}^z$ is 1 if there is money in the real wallet, and is 0 otherwise.
Constraints (\ref{eq:DFM-virtual-recharge}) update the virtual wallet balance $\mathbf{x}_t$ at each timestep $t$; the constant daily virtual wallet recharge $X_d$ is included in the constraints corresponding to the first timestep of the day. Constraints (\ref{eq:DFM-virtual-enable}), (\ref{eq:DFM-virtual-disable}) ensure that the virtual enable signal for load $k$, $\mathbf{u}_{k,t}^x$, is 1 if $\mathbf{x}_t$ is greater than or equal to the threshold $\mathbf{x}_{k,d}$, and is 0 otherwise. 
Constraints \eqref{eq:actuation-1}, \eqref{eq:actuation-on} ensure that a load $k$ at time $t$ is OFF, i.e., actuation state $\mathbf{a}_{k,t} = 0$, if there is no demand, if the virtual enable signal is 0, or if the real enable signal of the current or next timestep is 0. Similarly, they also ensure that $\mathbf{a}_{k,t} = 1$ if there is demand, virtual enable signal is 1, and the real enable signals for the current and next timestep are 1.
Since $\mathbf{a}_{k,t}, \mathbf{u}_{k,t}^z, \mathbf{u}_{k,t}^x$ are binary variables this is a MILP problem. The problem is always feasible if constants $\epsilon$, $m$, and $M$ are chosen such that: $\epsilon$ is as small as possible (e.g., equal to tolerance of the solver), $m \leq -Z$, $M \geq Z$.

\subsection{Average Forecast Greedy (AFG) model}
The average forecast greedy (AFG) model is the new model we propose in this work. 
Rather than requiring detailed load forecasts, this model only assumes knowledge of the \emph{average} power demand $\bar{P}_{k,d}$ per load $k$ per day $d$, which can be obtained by forecasting total energy consumption per load per day and dividing it by $24\text{ }\mathrm{h}$.
The decision variables in the model are the \emph{durations} $\mathbf{s}_{k,d}$ for which a load $k$ is enabled on day $d$. 
Based on the optimal solution $\mathbf{s}_{k,d}^*$, we can algebraically compute the optimal daily virtual wallet recharge, $\mathbf{X_d}$, and thresholds per load $k$ per day $d$, $\mathbf{x}_{k,d}$. The AFG model is described by the optimization model \eqref{eq:AFG-objective}-\eqref{eq:AFG-max-duration}. 
The objective (\ref{eq:AFG-objective}) is to maximize the PSF. Constraint (\ref{eq:AFG-balance}) ensures that the total cost of energy usage for all loads remains within the initial wallet balance, while (\ref{eq:AFG-max-duration}) provides an upper bound $s_{k,d}^{max}$ for the enable duration of each load. This upper bound $s_{k,d}^{max} = 0$ if $\bar{P}_{k,d} = 0$, i.e., when there is no demand for the load on that day, and $s_{k,d}^{max} = 24\text{ }\mathrm{h}$ otherwise. 
\begin{subequations}
\label{eq:AFG}
\begin{align}
    &\max_{\mathbf{s}_{k,d}}~~ &&\mathrm{PSF} = 
    \sum \limits_k \gamma_k \frac{\sum \limits_d \mathbf{s}_{k,d} }{\sum \limits_d s_{k,d}^{max}}
    \label{eq:AFG-objective}\\
    &\text{~s.t.~~}&&\sum \limits_{k,d} \alpha \bar{P}_{k,d} \mathbf{s}_{k,d} < Z
    \label{eq:AFG-balance}\\
    &&&\mathbf{s}_{k,d} \leq s_{k,d}^{max} \quad \forall k \in \mathcal{K}, \forall d \in \mathcal{D}
    \label{eq:AFG-max-duration}
\end{align}
\end{subequations}

\noindent
\subsubsection{Solving the AFG model}
Since $\mathbf{s}_{k,d}$ are continuous variables, the AFG model \eqref{eq:AFG} is a version of a fractional Knapsack problem, which seeks to maximize the value of items in our knapsack while respecting an overall weight limit. This type of problem can be solved efficiently to optimality using a greedy approach \cite{dantzig1957discrete}. In our problem, the ``items'' to be chosen are enable durations $\mathbf{s}_{k,d}$ per load per day and the ``value" of each item is the respective objective function coefficient 
$b_{k} = \frac{\gamma_k}{ \sum \limits_d s_{k,d}^{max} }$. The ``weight" of each item is the cost of using the load per hour for that day, i.e., $w_{k,d} = \alpha \bar{P}_{k,d}$. To solve the problem, we first compute the ratios of benefit per unit weight, $r_{k,d} = b_{k}/w_{k,d}$. Note that if $w_{k,d} = 0$, i.e., if there is no demand for a load on a day, we assign $\mathbf{s}^{*}_{k,d} = 0$. We arrange the ratios $r_{k,d}$ in non-increasing order in a vector $\mathbf{r}$. Therefore, we have $\mathbf{r}^{(i)} \geq \mathbf{r}^{(i+1)}$, where $\mathbf{r}^{(i)}$ represents the $i$th element of the vector $\mathbf{r}$. 
Next, we arrange variables $\mathbf{s}_{k,d}$ in a vector $\mathbf{s}$, parameters $s_{k,d}^{max}$ in vector $\mathbf{s}^{max}$, and parameters $\bar{P}_{k,d}$ in vector $\bar{\mathbf{P}}_{k,d}$ in the same order as that of elements in $\mathbf{r}$, such that the $i$th element of $\mathbf{s}$, $\mathbf{s}^{max}$,  and $\bar{\mathbf{P}}_{k,d}$ map to the same load $k$ and day $d$ as the $i$th element of $\mathbf{r}$. 
Given these vectors, we implement the greedy solution approach summarized in Algorithm \ref{alg:greedy}. Starting from the first element in $\mathbf{r}$, we assign the maximum duration $\mathbf{s}^{max (i)}$ to each variable $\mathbf{s}^{(i)}$, until we reach the element $\mathbf{s}^{(i^{'})}$ which leads to a violation of the wallet balance constraint (\ref{eq:AFG-balance}). For this \emph{marginal load} $\mathbf{s}^{(i^{'})}$, we assign a fractional value equal to the leftover balance divided by the cost of using the load per hour, which is typically less than $s_{k,d}^{max}$ (i.e., we enable the load only for a fraction of the day). For all elements after the $i^{'}$th element, i.e., those with a lower $r_{k,d}$ value, we set $\mathbf{s}^{(i)}$ to zero. This gives us the optimal enable durations $\mathbf{s}_{k,d}^*$ for each load $k$ and day $d$. Note that at most one load $k$ in one day $d$ will have a fractional value, i.e., there is only one marginal load.

\begin{algorithm}
\caption{Greedy approach to solve \eqref{eq:AFG-objective}-\eqref{eq:AFG-max-duration}} \label{alg:greedy}
\begin{algorithmic}
\STATE sort $\mathbf{r}$ in non-increasing order 
\STATE $\mathbf{s}^{(i)} \gets 0 \quad \forall i $
\STATE $i \gets 0$
\WHILE{$\sum \limits_{i} \alpha \bar{\mathbf{P}}^{(i)} \mathbf{s}^{(i)} < Z$}
\STATE{$\mathbf{s}^{(i)} \gets \mathbf{s}^{max(i)}$}
\STATE $i \gets i + 1$
\ENDWHILE
\STATE $i^{'} \gets i$
\vspace{1mm}
\STATE $\mathbf{s}^{(i^{'})} = \frac{Z - \sum \limits_{i} \alpha \bar{\mathbf{P}}^{(i)}\mathbf{s}^{(i)}}{\alpha \bar{\mathbf{P}}^{(i^{'})}}$
\end{algorithmic}
\end{algorithm}

\noindent
\subsubsection{Computing virtual wallet recharge} The daily virtual wallet recharge $\mathbf{X}_d$ is calculated as the amount of recharge needed to support the optimal enable durations $\mathbf{s}_{k,d}^*$, i.e., $$\mathbf{X}_d = \sum \limits_k \alpha\mathbf{s}_{k,d}^*\bar{P}_{k,d}, \forall d \in \mathcal{D}$$ 
The virtual wallet is recharged by $\mathbf{X}_d$ at the beginning of the day with no further recharge during the day, and the virtual wallet balance thus reduces as the loads consume energy. Once the balance goes below threshold $\mathbf{x}_{k,d}$, the corresponding load $k$ is disabled and not enabled again on that day. 
\subsubsection{Computing thresholds} Algorithm \ref{alg:compute_thresholds} presents the method to compute the thresholds $\mathbf{x}_{k,d}$. To compute the thresholds $\mathbf{x}_{k,d}$ for a given day $d$ for a given load $k$, we distinguish three cases. First, if $\mathbf{s}_{k,d}^* = 0$ (i.e., the load is disabled for the whole day), then the threshold $\mathbf{x}_{k,d}$ is assigned a value slightly larger than the virtual wallet recharge for the day, $\mathbf{x}_{k,d}=\mathbf{X}_d + \epsilon$, where $\epsilon = 10^{-4}$. The virtual wallet balance is expected to never exceed the initial charge $\mathbf{X}_d$ and by setting the threshold higher than this value, the model expects that it will not be enabled. Second, if $\mathbf{s}_{k,d}^* = s_{k,d}^{max}$ (i.e., the load should be enabled for the whole day), then the threshold $\mathbf{x}_{k,d}$ is set to zero. This ensures that the virtual wallet balance never goes negative. 
Finally, we consider the threshold for the marginal load $\mathbf{s}_{k,d}^*$ (corresponding to $\mathbf{s}^{(i^{'})}$) that has $0\leq\mathbf{s}_{k,d}^*\leq s_{k,d}^{max}$.
The threshold of this load $\mathbf{x}_{k,d}$ should be chosen such that 
the virtual wallet balance, which is reducing from $\mathbf{X}_d$ in the beginning of the day, reaches the threshold $\mathbf{x}_{k,d}$ after $\mathbf{s}_{k,d}^*$ hours. This can be expressed as 
$$\mathbf{x}_{k,d} = \mathbf{X}_d - \alpha \mathbf{s}_{k,d}^*\sum\limits_{n\in \mathcal{S}_d^{en}}\bar{P}_{n,d},$$
where the second term is the cost of running all enabled loads on the day, $\mathcal{S}_d^{en} \equiv \{n:\mathbf{s}_{n,d}^{*} > 0\}$, for the duration $\mathbf{s}_{k,d}^*$. The overall time complexity of the AFG model is very low and depends mainly on the type of sorting algorithm used. For example, if merge sort is used, it will be $\mathcal{O}(n\log{}n)$.

\begin{algorithm}
\caption{Computing thresholds} \label{alg:compute_thresholds}
\begin{algorithmic}
\FOR{$d \in \mathcal{D}$}
\FOR{$k \in \mathcal{K}$}
\IF{$\mathbf{s}_{k,d}^{*} == 0$}
\STATE{$\mathbf{x}_{k,d} \gets \mathbf{X}_d + \epsilon$}
\ELSIF{$\mathbf{s}_{k,d}^{*} == s_{k,d}^{max}$}
\STATE{$\mathbf{x}_{k,d} \gets 0$}
\ELSE
\STATE{$\mathbf{x}_{k,d} \gets \mathbf{X}_d - \alpha \mathbf{s}_{k,d}^*\sum\limits_{n\in \mathcal{S}_d^{en}}\bar{P}_{n,d}$}
\STATE{where $\mathcal{S}_d^{en} \equiv \{n:\mathbf{s}_{n,d}^{*} > 0 \}$}
\ENDIF
\ENDFOR
\ENDFOR
\end{algorithmic}
\end{algorithm}

\subsection{Optimal Benchmark MILP (OBM) model}
We now describe a model more closely aligned with traditional home energy management systems proposed in literature. Rather than relying on thresholds, this model directly decides the activation state of each load at each time step. If given perfect forecasts of desired consumption, it produces an optimal schedule with the highest possible PSF for a given budget. Therefore, we refer to this as the optimal benchmark MILP (OBM) model, which is given by 
\begin{subequations}
\begin{align}
        &\max_{\mathbf{a}_{k,t}} &&\mathrm{PSF} = \sum\limits_k \gamma_k \frac{\sum\limits_t \mathbf{a}_{k,t}}{\sum\limits_t d_{k,t}}
    \label{eq:OBM-objective-OBM} \\
    &\text{~s.t.}&&\sum\limits_{k,t}\mathbf{a}_{k,t}\alpha P_{k,t} \Delta T < Z
    \label{eq:OBM-balance} 
    \\
    &&&\mathbf{a}_{k,t} \leq d_{k,t} \quad \forall k \in \mathcal{K}, \forall t \in \mathcal{T}
    \label{eq:OBM-a-leq-d}
\end{align}
\end{subequations}
The model inputs are the demand forecasts $P_{k,t}$ per load $k$ at each timestep $t$, 
and it determines the actuation state $\mathbf{a}_{k,t}$ per load $k$ at time $t$, so as to maximize the PSF as defined by (\ref{eq:OBM-objective-OBM}). Constraint (\ref{eq:OBM-balance}) ensures that the energy usage is within the initial real wallet balance $Z$ and constraint (\ref{eq:OBM-a-leq-d}) ensures that a load is actuated only when there is demand. Since $\mathbf{a}_{k,t}$ are binary variables, it is a MILP problem.

\section{Qualitative Comparison - Computation, Communication, and Information Requirements}
\label{section:hardware-communication-requirements}

In this section, we compare the computation, communication, and information requirements of the proposed AFG model with the DFM and OBM models. Ideally, we want a method that limits the need for expensive local hardware, while also minimizing communication needs. 
We frame our discussion in the context of two possible modes of implementation, \emph{purely local}, where all computations are performed using local in-home hardware, or \emph{mixed mode}, where the optimization problems are solved in the cloud on a remote server and the setpoints are communicated to the local hardware.
\vspace{-0.8mm}
\subsection{Computational requirements}
OBM and DFM are MILP problems with a large number of variables, which are generally hard to solve. 
In our test cases, the DFM problem can take up to 50 min to converge (with 30 day optimization horizon, 4 loads) on a system with a state-of-the-art commercial MILP solver. The computational effort required to solve the problem indicates that a local implementation of the problem may be impractical or expensive. 
In comparison, AFG is an LP problem that can be solved to optimality with the greedy approach outlined above, making it quick and easy to solve even with simpler hardware and without optimization solvers. Therefore, it can be implemented purely locally on existing hardware (e.g., the in-home displays typically provided to prepaid customers \cite{2010_epri}).

\subsection{Load information requirements}
Load forecasting is challenging, but can be achieved using statistical and machine learning based methods \cite{li2021short}. The DFM and OBM models require load forecasts with a $15\text{ } \mathrm{min}$ granularity for each load. Creating accurate forecasts with such granularity is a very challenging task and 
would require customers to either
share granular historical load data with the remote server, or generate the forecasts locally and communicate them. Either option may require significant communication bandwidth or local computational power. In comparison, the AFG model only requires information regarding the total expected energy demand for each load per day, from which the average power demand can be computed. This lower granularity forecast would be significantly easier to generate, whether it is done locally or remotely. Furthermore, we would only need to communicate one value per load per day. Avoiding the sharing of high fidelity load forecasts is also more desirable from a privacy and cybersecurity perspective.

\subsection{Communication of load activation information}
In the \emph{mixed mode} of implementation, the optimization problems are solved on a remote server and we need to communicate the resulting control setpoints (i.e., load activation information) to the user. For DFM, only a single threshold per load per day needs to be communicated from the remote server to the local hardware, while AFG requires that the thresholds and the virtual wallet recharge per day be communicated. For a household with $|\mathcal{K}|$ loads, this means communicating $|\mathcal{K}|$ 
and $|\mathcal{K}|+1$ floating-point numbers per day respectively.
For OBM, the actuation states per load per timestep need to be communicated. This amounts to $|\mathcal{K}||\mathcal{T}|$ binary numbers per day where $|\mathcal{T}|$ is the number of timesteps in a day. This is a much larger number than the number of setpoints to be communicated for implementing AFG or DFM. Note that for the AFG model, the communication of control setpoints could be entirely avoided through a purely local implementation. 
This also proves to be useful in case the model has to be re-run with new load priority factors. For example, in case AFG notifies the user that an uninterruptible load such as a dishwasher has to be turned off in the middle of its operating cycle,
the user can choose to assign a relatively higher priority factor to the load and re-run the model locally for a new set of thresholds.

\begin{table}[]
    \caption{\small Qualitative comparison of models}
    \begin{tabular}{L{0.2\columnwidth} L{0.2\columnwidth} L{0.2\columnwidth} L{0.2\columnwidth}}
\hline
 & \textbf{OBM} & \textbf{DFM} & \textbf{AFG}\\
\hline
Problem type & MILP & MILP & LP\\
\hline
Variables per day & $|\mathcal{K}||\mathcal{T}|$  & $|\mathcal{T}|(3|\mathcal{K}|$\newline$ +2) +|\mathcal{K}|$
& $|\mathcal{K}| + 1$\\
\hline
Load forecast information & per timestep per load & per timestep per load & average per day per load \\
\hline 
Control setpoints per day & 
$|\mathcal{K}||\mathcal{T}|$ binary numbers &
$|\mathcal{K}|$ floating-point numbers & 
$|\mathcal{K}| + 1$ floating-point numbers \\
\hline
Mode of implementation & mixed & mixed & purely local \\
\hline
    \end{tabular}
    \label{tab:comparison-models}
\vspace{-5mm}
\end{table}

An overview of the comparison is summarized in Table \ref{tab:comparison-models}.
We observe that the AFG model has several significant advantages from an implementation perspective. It is the least computationally expensive method, making it suitable for a purely local implementation. If implemented in mixed mode, it communicates only low-dimensional information regarding loads and control setpoints, thus 
reducing 
privacy and cyber-security concerns.

\section{Quantitative Comparison - Case Study}
\label{section:case-studies}
We next compare the AFG model with the DFM and OBM models in a quantitative case study based on 
real-life energy usage data. 

\subsection{Case study setup}
\label{section:case-study-setup} 
We use energy usage data of one household from the Pecan Street Dataset \cite{noauthor_dataport_nodate} as load data and ``forecasts". We consider four loads, namely a refrigerator, air compressor, microwave, and washing machine in that priority order. The priority factors of the four loads are $\gamma = 0.48, 0.24, 0.16, 0.12$ respectively. The cost of electricity is assumed to be $\alpha = 0.16\text{ }\mathrm{\$/kWh}$. 
Each optimization model solves a problem to obtain setpoints, for a duration of 30 days. These setpoints are then used as input to a numerical simulation for the same 30 days. This process is repeated for three months of data. The Julia programming language (v1.6) is used with the JuMP package \cite{dunning2017jump} and  the Gurobi solver \cite{gurobi} for implementing the models on an Intel CPU @3.2GHz machine with 16GB memory.

We compare the priority service factor (PSF) from the numerical simulation for AFG, DFM, OBM, and a simple baseline (BSL) of unrationed energy use, which lets a user satisfy all energy demand until the wallet balance is zero. We consider a \emph{perfect} demand forecast, as well as an \emph{imperfect} forecast where the order of days is shuffled. Within each type, we consider two levels of information granularity, a \emph{detailed} forecast with $15\text{ } \textrm{min}$ power demand per load and \emph{limited} forecast with only daily average power demand per load. To reflect the reality of low income customers who may not 
have
enough money to cover their desired energy demand, we assume initial wallet balances that are enough to supply 70\% to 90\% total energy demand. The cases and results are described in more detail below.

\subsection{Perfect forecast}
\label{section:perfect-forecast}
We first assume perfect forecasts of load demand, and investigate the detailed and limited information cases. 

\subsubsection{Detailed information}
In this set of experiments, models know the true load demand at each $15\text{ } \mathrm{min}$ timestep in the case of OBM and DFM and the true average demand per load per day in the case of AFG. 
With this perfect detailed information, the OBM provides a truly optimal solution since it has complete information about load demand and can decide which load to actuate at every timestep. The DFM also has information about load demand at each timestep but can only determine a threshold for enabling each load per day, thus we expect DFM to have a lower PSF than OBM. Since AFG only uses daily average power demand per load as input, we expect AFG to have a lower PSF than both DFM and OBM. 
In these experiments, we seek to assess the performance drop of AFG and DFM relative to OBM. 

Figure \ref{fig:perfect-limited}(a) shows improvement in PSF over the baseline for month A, while the results for all months A, B, and C are shown in Table \ref{tab:perfect-forcast}. Note that the DFM values are from the optimization model (and not numerical simulation) because of numerical precision issues.
We observe that OBM has the highest improvement for all recharge amounts and months. The performance improvement of DFM is $0-2\ \%$pt. lower than OBM depending on the month and recharge level, indicating that there is some performance drop due to only having the thresholds as a control variable. The AFG model improves performance relative to the baseline case across all months, despite having significantly lower granularity of demand information. The PSF improvement is $2-6\ \%$pt. lower than OBM and $0.5-5\ \%$pt. lower than DFM, indicating that detailed load forecasts are important to achieve high performance.

\subsubsection{Limited information}
To further assess the impact of limited information, we run experiments where we provide all models with information about the average demand per load per day (i.e., the same information provided to AFG in the previous case). This is a more realistic case, with an easier to obtain load forecast. Further, in this case, the OBM and DFM have no obvious advantage over the AFG model.

\begin{figure}
\centering
\subfloat[\small Detailed information]
{\includegraphics[width=0.49\linewidth]{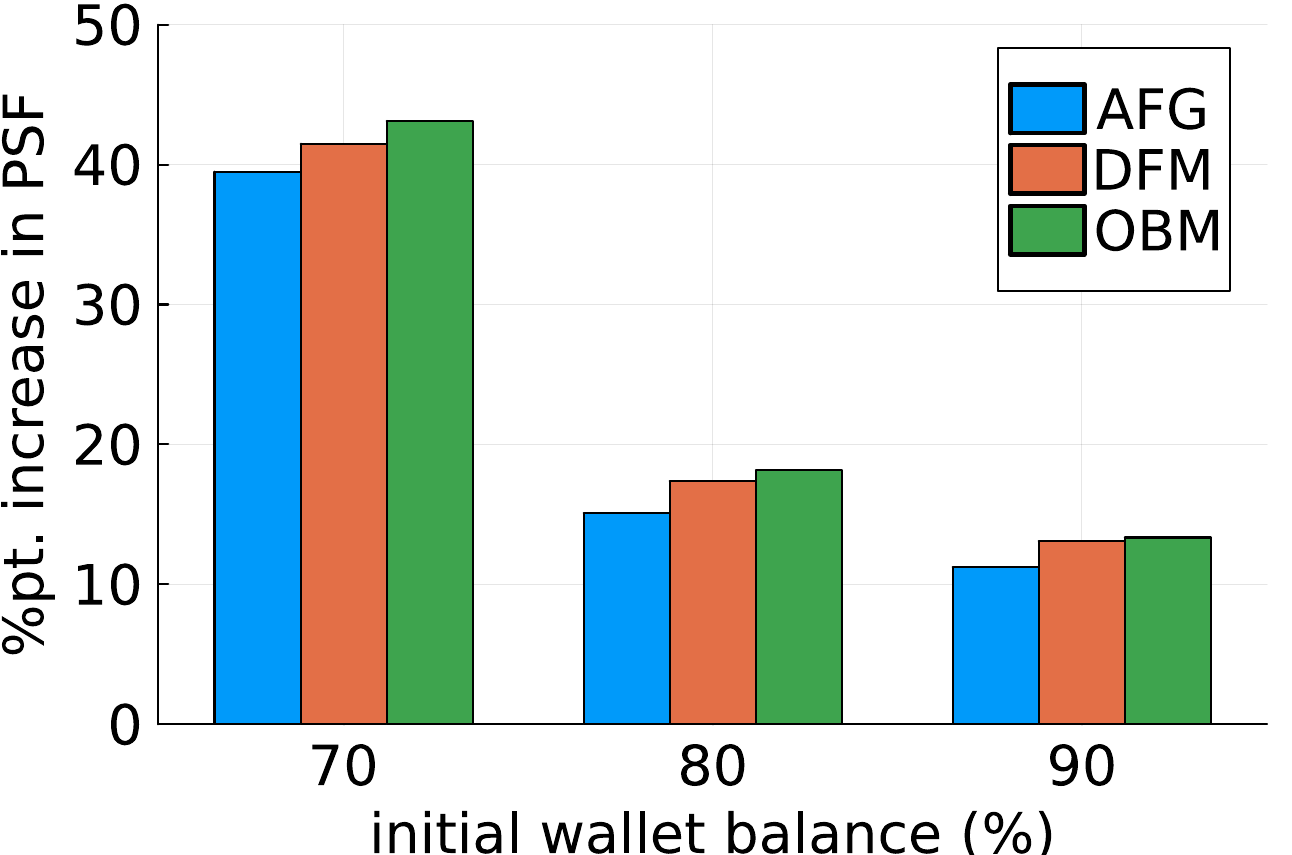}}
\hfil
\subfloat[\small Limited information]
{\includegraphics[width=0.49\linewidth]{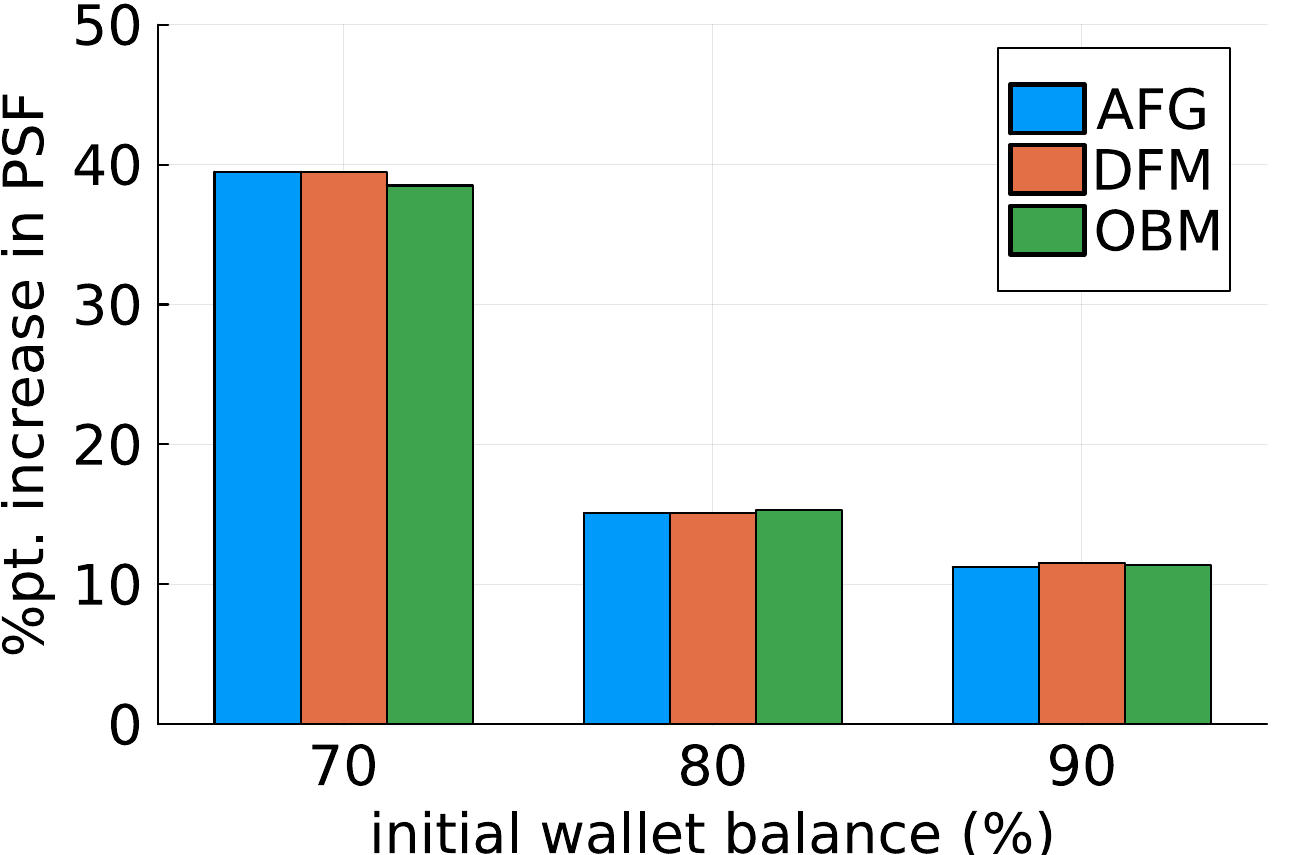}}
\vspace{-1mm}
\caption{\small Perfect forecast: Percentage point (\%pt.) improvement in priority service factor (PSF) over baseline for the proposed AFG model (blue) and the benchmark models DFM (red) and OBM (green)}
\label{fig:perfect-limited}
\end{figure}

Figure \ref{fig:perfect-limited}(b) shows the improvement in PSF over the baseline for each model for month A, with similar numbers provided for months A, B, and C in Table \ref{tab:perfect-forcast}. AFG and DFM perform within $1\ \%$pt. of each other in all cases, while the OBM model performs similar to the others (within $1\ \%$pt. in most cases) or up to $4\ \%$pt. worse in the high initial balance cases in month B. This confirms that all models perform similarly in the setting where they all are given limited information. 

\begin{table}
\caption{\small Perfect forecast: Percentage point improvement in priority service factor (PSF) over baseline for the proposed AFG model and the benchmarks DFM and OBM}
    \begin{tabular}{c|c|ccc|ccc}
    \multirow{2}{*}{data} & \multirow{2}{*}{balance} & \multicolumn{3}{c|}{perfect detailed} & \multicolumn{3}{c}{perfect limited} \\
         & & AFG  & DFM & OBM & AFG & DFM & OBM\\
         \hline
        
        \multirow{3}{*}{A}& 70\% & 39.5 & 41.5 & 43.1 & 39.5 & 39.5 & 38.5\\
        & 80\% & 15.1 & 17.4 & 18.2 & 15.1 & 15.1 & 15.3\\
        & 90\% & 11.2 & 13.1 & 13.3 & 11.2 & 11.5 & 11.4\\
        \hline
        \multirow{3}{*}{B} & 70\% &  13.7 & 14.6 & 16.8 & 13.7 & 13.6 & 12.9\\
        & 80\% & 10.6 & 11.3 & 13.4 & 10.6 & 10.5 & 6.97\\
        & 90\% & 8.35 & 9.01 & 10.5 & 8.35 & 8.33 & 5.99\\
        \hline
        \multirow{3}{*}{C}& 70\% & 8.32 & 13.2 & 14.0 &  8.32 & 8.30 & 8.76\\
        & 80\% & 3.63 & 8.64 & 9.02 & 3.63 & 4.24 & 4.21\\
        & 90\% & 2.16 & 5.15 & 5.18 & 2.16 & 2.94 & 2.94
    \end{tabular}
    \label{tab:perfect-forcast}
    \vspace{-4mm}
\end{table}

\begin{table*}[]
\centering
\caption{\small Imperfect forecast: PSF in percent and improvement in PSF over baseline BSL in percentage points in parentheses for the proposed model AFG, the benchmarks DFM and OBM, and PSF and disconnection duration (\# days) for the BSL model}
\begin{tabular}{c|l|lll|lll|ll}
\multirow{2}{*}{month} & \multirow{2}{*}{balance} & \multicolumn{3}{c|}{imperfect   detailed}      & \multicolumn{3}{c|}{imperfect limited}     &  & \multirow{2}{*}{\# days} \\
                       &                          & AFG            & DFM            & OBM            & AFG            & DFM            & OBM & BSL                                          \\
                       \hline
\multirow{3}{*}{A}     & 70\%                     & 80.6 (29.2)    & 77.8 (26.3)    & 33.7 (-17.8)   & 80.6 (29.2)    & 79.1 (27.6)    & 70.0 (18.5)    & 51.5                                     & 17 \\
                       & 80\%                     & 80.6 (1.95)    & 77.5 (-1.15)   & 34.0 (-44.7)   & 80.6 (1.95)    & 78.1 (-0.63)   & 70.0 (-8.69)   & 78.7                                     & 7 \\
                       & 90\%                     & 80.6 (-4.69)   & 73.5 (-11.8)   & 34.2 (-51.1)   & 80.6 (-4.69)   & 80.6 (-4.76)   & 70.0 (-15.3)   & 85.3                                     & 5 \\
                       \hline
\multirow{3}{*}{B}     & 70\%                     & 71.0 (3.01)    & 73.1 (5.08)    & 33.8 (-34.2)   & 71.0 (3.01)    & 76.2 (8.22)    & 61.9 (-6.07)   & 68.0                                     & 9                                                            \\
                       & 80\%                     & 71.0 (-6.06)   & 80.5 (3.41)    & 35.1 (-41.9)   & 71.0 (-6.06)   & 76.3 (-0.74)   & 61.9 (-15.1)   & 77.1                                     & 6                                                            \\
                       & 90\%                     & 72.9 (-12.4)   & 81.2 (-4.08)   & 36.9 (-48.4)   & 72.9 (-12.4)   & 78.6 (-6.69)   & 69.6 (-15.7)   & 85.3                                     & 4                                                            \\
                       \hline
\multirow{3}{*}{C}     & 70\%                     & 78.2 (-1.25)   & 78.7 (-0.82)   & 33.0 (-46.5)   & 78.2 (-1.25)   & 75.6 (-3.88)   & 72.8 (-6.69)   & 79.5                                     & 3                                                            \\
                       & 80\%                     & 88.4 (1.31)    & 86.3 (-0.80)   & 34.0 (-53.2)   & 88.4 (1.31)    & 85.8 (-1.30)   & 73.4 (-13.8)   & 87.1                                     & 2                                                            \\
                       & 90\%                     & 95.2 (2.26)    & 94.9 (1.96)    & 34.1 (-58.8)   & 95.2 (2.26)    & 95.2 (2.26)    & 73.4 (-19.6)   & 92.9                                     & 1 \\
\end{tabular}
\label{tab:imperfect-forecast}
\vspace{-4mm}
\end{table*}

\subsection{Imperfect forecast}
\label{section:imperfect-forecast}
Next, we assess performance under imperfect forecast information, generated by randomly shuffling the order of days. 
The order of shuffling is preserved across models. 
\subsubsection{Detailed information} 
First, we provide DFM and OBM with (imperfect) 15 min demand information 
whereas AFG is provided with (imperfect) average power demand information per day. 
Figure \ref{fig:imperfect-imperfect-limited}(a) shows the PSF for each model for month A, while Table \ref{tab:imperfect-forecast} shows the results for months A, B, and C. We observe that in month A, the AFG model outperforms the OBM and DFM models, while when considering all the months in Table \ref{tab:imperfect-forecast}, we see that PSF of AFG was sometimes lower than that of DFM. Therefore, the relative performance of the two models depends on the specific data used. The OBM model has the lowest PSF (up to $61\ \%$pt. lower than AFG) across all months and recharge levels, because it activates loads at specific times  which do not coincide with the true demand. 
It is also worth noting that the PSF of the baseline BSL (i.e., no control) is comparable to and sometimes higher than that of the other models in the imperfect information case. However, the user experiences a disconnection in each case with the BSL and may require the payment of reconnection fees. Further, the baseline case leaves the user with no access to power for a prolonged period of time after the disconnection, e.g., 
for month A the disconnection time is
17, 7, and 5 days
with
70\%, 80\%, and 90\% initial wallet balance, respectively. 
Furthermore, it is also likely that a more realistic implementation of the AFG, DFM, and OBM models, where setpoint optimization would be rerun frequently (e.g., daily) with information from actual usage in the previous day, would help close the performance gap with the baseline BSL model.

\subsubsection{Limited information}
\begin{figure}[htbp]
\centering
\subfloat[{\small Detailed information}]
{\includegraphics[width=0.49\linewidth]{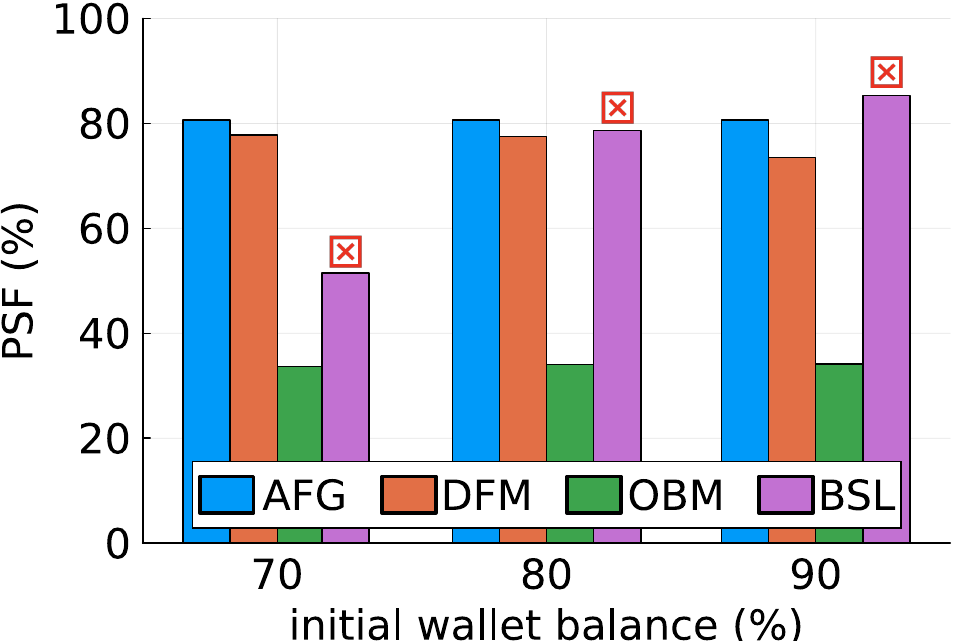}}
\hfil
\subfloat[{\small Limited information}]
{\includegraphics[width=0.49\linewidth]{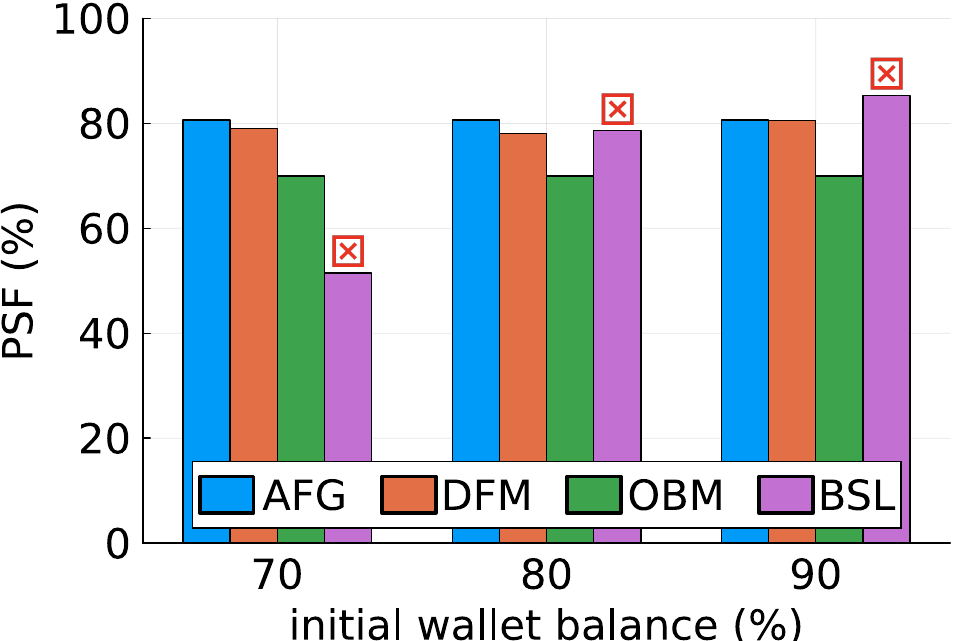}}
\vspace{-1mm}
\caption{\small Imperfect forecast: Priority service factor (PSF) and customer disconnections (indicated by `x' mark) for the proposed model AFG (blue), the benchmark models DFM (red), OBM (green), and the baseline BSL model (purple)}
\label{fig:imperfect-imperfect-limited}
\end{figure}

\noindent
Next, we provide all models only average power demand data for the shuffled days (same as what was provided to AFG in the previous case) and results are shown in Figure \ref{fig:imperfect-imperfect-limited}(b) for month A and Table \ref{tab:imperfect-forecast} for months A, B, and C.
Similar to the imperfect-detailed case, PSF of AFG is higher than that of OBM (up to $22\ \%$pt.)  and the relative performance of AFG and DFM is observed to be dependent on the specific data used as seen in Table \ref{tab:imperfect-forecast}. 
It can also be seen that the PSF of OBM considerably increases. This is because the forecast contains demand equaling the daily average power demand for the entire duration of 24 hours every day, which  
may have more overlap with the actual demand than in the previous case. 

These results indicate that when all models are provided with perfect-limited information, the AFG model performs on par with DFM and OBM. With imperfect-limited information, AFG performs on par with DFM in most cases and outperforms OBM. The case study highlights that the proposed AFG model achieves comparable or improved performance compared to the benchmark models, despite being computationally simpler.

\vspace{-3mm}
\section{Conclusion}
This study presents a linear threshold-based energy rationing model for prepaid electricity customers, which needs only daily average demand forecasts and can be solved to optimality using a simple greedy strategy. We compare this model with two mixed-integer linear programming (MILP) models which require demand forecasts at each $15\text{ }\mathrm{min}$ timestep in case studies with perfect, limited, and imperfect forecast information. The proposed linear model has comparable or improved performance compared with that of the MILP models, while being computationally inexpensive. Therefore, it can be implemented on inexpensive local in-home hardware, which is desirable from a privacy and cybersecurity perspective.

Some avenues for future work include modeling user noncompliance to load enable/disable signals, determining a relationship between forecast error and model performance, formulating the optimization model to incorporate uncertainty in forecasts, and studying the effects of delays in computation and communication of optimal setpoints for energy rationing. We would also like to deploy this method in a field experiment. 

\bibliographystyle{IEEEtran}
\bibliography{myBibliography}

\end{document}